\documentclass[aps,prd,preprintnumbers,nofootinbib]{revtex4}
\usepackage[utf8]{inputenc}
\usepackage{bm}
\usepackage{latexsym}
\usepackage{dcolumn}
\usepackage{amsmath,amsfonts,amssymb}
\usepackage{graphicx,epsfig}
\usepackage{psfrag}
\usepackage{amsthm}
\usepackage[pdftex]{hyperref}
\usepackage[colorinlistoftodos]{todonotes}

\def\be{\begin{eqnarray}}
\def\ee{\end{eqnarray}}

\renewcommand\t{\tilde}

\begin{document}

\title{Constraints and Horizons for de Sitter with Extra Dimensions}
\author{Saurya Das}\email{saurya.das@uleth.ca}
\affiliation{Theoretical Physics Group, \\
Department of Physics and Astronomy,\\
University of Lethbridge, 4401 University Drive,\\
Lethbridge, Alberta T1K 3M4, Canada}
\author{S. Shajidul Haque}\email{shajid.haque@uwindsor.ca}
\affiliation{Department of Physics,\\
University of Windsor,\\
Windsor, Ontario N9A 0C5, Canada}
\author{Bret Underwood}\email{bret.underwood@plu.edu}
\affiliation{Department of Physics,\\
Pacific Lutheran University,\\
Tacoma, WA 98447}
\date{\today}

\vspace{-2cm}
\begin{abstract}

In order for spacetimes with static extra dimensions to have 4-dimensional de Sitter expansion they must have at least positive curvature, warping sourced by the 4-d expansion, or violate the null energy condition everywhere in the extra dimensions.
We show how this constraint arises from the null Raychaudhuri equation, and that it is independent of the matter content, the Einstein equations, and is true point-by-point in the extra dimensions (not integrated), setting it apart from other no-go theorems in the literature.
We present two previously known examples -- a Freund-Rubin compactification with bulk cosmological constant, and a Randall-Sundrum model where the mismatch between the brane and bulk cosmological constants sources non-trivial warping -- which evade the constraint, and  discuss the implications for flux compactifications.
We also show that any spacetime with static compact extra dimensions and 4-dimensional de Sitter expansion has an apparent horizon and antitrapped region in the extra dimensions, which may have interesting implications for the dynamics of bulk fields in cosmology.

\end{abstract}

\maketitle

\section{Introduction}

Extra dimensions are a common ingredient of high energy physics theories beyond the Standard Model.
An essential feature of any theory with $D>4$ spacetime dimensions is that it must be able to reproduce cosmological observations in 4-dimensions, including periods of de Sitter (dS) or near-dS accelerating expansion.
While it is relatively straightforward to obtain dS space classically in a purely 4-dimensional theory (by introducing a 4-dimensional cosmological constant), it is surprisingly difficult to construct models with 4-dimensional dS space in the presence of other static (non-expanding) extra dimensions.

In particular, there are a number of ``no-go'' theorems that detail how the matter content of the higher dimensional theory must violate an energy condition in the extra dimensions in order to obtain 4-dimensional dS \cite{Gibbons:2003gb,MN,Wesley:2008fg,Steinhardt:2008nk,Postma,Douglas:2010rt,Dasgupta:2014pma}. Some separate related lines of investigation have been concerned with whether dS space can be explicitly realized in compactifications of string theory \cite{Silverstein:2008sg,Hertzberg:2007wc,Silverstein:2007ac,Haque:2008jz, Danielsson:2009ff,Danielsson:2011au,Wrase:2010ew,Andriot:2016xvq,Andriot:2018ept,Cordova,Cribiori:2019clo,Andriot:2019wrs,Heckman:2018mxl,Heckman:2019dsj}, see also \cite{Kutasov:2015eba,Green:2011cn,Sethi:2017phn}, as well as the consistency of the model known as KKLT \cite{KKLT} as in \cite{Carta:2019rhx,Gautason:2019jwq,Hamada:2019ack,Bena:2009xk,Blaback:2012nf,Gautason:2013zw,Moritz:2017xto,Danielsson:2018ztv}.
Finally, recent speculation has suggested that 4-dimensional dS space may be inconsistent with quantum gravity \cite{Obied:2018sgi,Agrawal:2018own, Ooguri:2018wrx,Akrami:2018ylq}.

The approaches described above make a number of apparently reasonable-looking 
assumptions in order to reach their conclusions.
Some of these assumptions, such as integrating out the extra dimensions and working in the 4-dimensional effective field theory, make it difficult to see how the solution will behave in higher dimensions.
Other assumptions, such as the existence of certain matter content, the validity of the Einstein equations, and the assumption that the extra dimensions have no boundary (so that integrals of total derivatives over the extra dimensions vanish), restrict the generality of the conclusions.
Further, many of the energy conditions in the ``no-go'' theorems are violated by well-known ingredients in string theory, such as orientifold planes. This can make it difficult to draw strong conclusions about the possibility of realizing 4-dimensional dS space in models that contain these ingredients.

In this manuscript, we will explore a different way of generating constraints on theories with 4-dimensional dS and extra dimensions.
Our approach will center on the null Raychaudhuri equation in $D>4$ dimensions. The power of the null Raychaudhuri equation is manifold:
it is a purely geometric identity, so it only requires the assumption of a metric ansatz, and is thus independent of the form of the Einstein equations and matter content of the model in consideration.
It is also a local equation, in that it applies at each point in the extra dimensions and is not integrated over the internal space (so that there are no assumptions about boundaries or boundedness).
Finally, it leads to a condition that in order to have 4-dimensional dS the model must violate the null energy condition (NEC) and/or the extra dimensions have positive curvature.
The NEC is satisfied by a large set of matter content, so this will further tighten the constraints on models with extra dimensions.

We will then show using the expansion scalar of null congruences that models which evade this constraint and have 4-dimensional dS with static extra dimensions necessarily possess an apparent horizon in the extra dimensions, with a corresponding anti-trapped region.
The apparent horizon is due to the shear from the expansion in the non-compact dimensions, and may have interesting implications for the cosmological dynamics of fields present in the bulk of the extra dimensions.
As with the previous result, this conclusion arises in a purely geometric way, and will only depend on the existence of a metric with 4-dimensional dS space and static extra dimensions.

The manuscript is organized as follows. In Section \ref{sec:NullRay}, we use the null Raychaudhuri equation with a generic metric ansatz to construct a constraint that models with extra dimensions must satisfy in order to realize 4-dimensional dS.
We will explore the relationship of this constraint to previous ``no-go'' theorems, and demonstrate that our constraint is significantly stronger.
In particular, we will show that matter sources or geometries (such as a bulk cosmological constant or negative curvature) that evade previous no-go's and are thought to be important ingredients in building putative dS solutions are, however, ruled out by our constraint.
We then illustrate two simple examples which evade our constraints: a 6-dimensional Freund-Rubin compactification of 2-form flux on a sphere with a bulk cosmological constant and a Randall-Sundrum (RS) model in which the brane and bulk cosmological constants are ``de-tuned.''
We finish this section with some comments on the role that the warp factor plays in the constraint for flux compactification models, as well as the possibility of higher curvature corrections playing an important role.
In Section \ref{sec:Horizon} we demonstrate that any theory with static extra dimensions and 4-dimensional dS space will have two apparent horizons and an antitrapped region in the extra dimensions. We conclude by briefly commenting on the possible implications of the horizon.
We have relegated some details to the Appendix.

\section{Satisfying the Null Raychaudhuri Equation}
\label{sec:NullRay}

In order for spacetimes with extra dimensions to make contact with cosmological observations, we need to allow for the 3 observed large spatial dimensions to be expanding. Further, current early- and late-time cosmological models incorporate a period of near-de Sitter (dS) accelerating expansion, and the possibility of a non-zero 4-dimensional cosmological constant poses interesting theoretical challenges on model building.

Based on these reasons, we will take as our metric a generic $D$-dimensional warped product of a 4-dimensional expanding spacetime with an $n$-dimensional manifold of extra dimensions $z^m = (\chi,y^m)$:
\be
ds^2 &=& \Omega^2(\chi,y^m) \left[\hat g_{\mu\nu} dx^\mu dx^\nu + \bar g_{mn} dz^m dz^n\right] \nonumber \\
&=& \Omega^2(\chi,y^m)\left[-dt^2 + a^2(t) \delta_{ij} dx^i dx^j + d\chi^2 + f^2(\chi)\, \t g_{mn}(y^m) dy^m dy^n\right]\, ,
\label{exMetric}
\ee
where we singled out a ``radial'' direction $\chi$, and assumed that the $\chi$-dependence of the rest of the metric can be factorized\footnote{We will relax both of these assumptions later, and find that the general conclusions remain the same.}. The warp factor $\Omega(\chi,y^m)$ can in general depend on any of the internal coordinates $(\chi,y^m)$.
For our analysis the extra dimensions need not be compact; they could be infinite in extent, or even bounded by branes. Importantly, our analysis will be strictly local in nature and thus will not rely on compactness.
The metric (\ref{exMetric}) is related to more commonly used metrics in string cosmology compactifications such as \cite{GKP}, 
including
\be
ds^2 = e^{2A(r,y^m)} \left(-dt^2 + a^2(t) \delta_{ij} dx^i dx^j\right) + e^{-2 A(r,y^m)} \left(dr^2 + f^2(r) \t g_{mn}(y^m) dy^m dy^n\right)\, ,
\ee
which is related to (\ref{exMetric}) by a conformal rescaling of the internal metric, a redefinition of the radial coordinate $\chi \rightarrow \chi(r)$, and a redefinition of the warp factor $e^A = \Omega$. However, we will find the form (\ref{exMetric}) to be
more convenient for our analysis.
Note that we have chosen (\ref{exMetric}) to take the form of a warped product, with no cross-terms $g_{t\chi},g_{tm}$ between the time and the internal coordinates. While this makes sense for 4-dimensional dS space (because the symmetries of dS would forbid such terms), it is possible such terms would be necessary for more generic cosmologies, such as considered in \cite{Wesley:2008fg,Steinhardt:2008nk,Postma}.

An interesting question is: what are the required conditions for the metric (\ref{exMetric}) to support accelerated 4-dimensional expansion, i.e.~a dS or near-dS space? 
Interestingly, this question can be answered in a purely geometric and model-independent way, without reference to the Einstein equations or explicit matter content, by using the Raychaudhuri equation.
Given a congruence of affine null tangent vectors $N^M$, these null rays must satisfy the geometric identity\footnote{We have set the twist tensor to zero $\omega_{AB} = 0$, reflecting the fact that our null vectors will be hypersurface orthongonal.}
\be
\frac{d\theta}{d\lambda} = -\frac{1}{D-2} \theta^2 - \sigma_{MN} \sigma^{MN} - R_{MN} N^M N^N\, ,
\label{NullRayEq}
\ee
known as the Raychaudhuri equation, where 
\be
\theta \equiv \frac{1}{\sqrt{-g_D}} \partial_A \left(\sqrt{-g_D} N^A\right)
\label{Expansion}
\ee
is known as the expansion and 
\be
\sigma_{MN} \equiv \frac{1}{2} \left(\nabla_M N_N + \nabla_N N_M\right) - \frac{1}{D-2} \hat h_{MN} \theta\, ,
\label{ShearTensor}
\ee
is known as the shear tensor, which is transverse to the null ray $\sigma_{MN} N^M = 0$,
and $\hat h_{MN}$ is the transverse metric, constructed with the help of a cross-normalized auxiliary vector $K^M$, $N^M K^N g_{MN} = -1$
\be
\hat h_{MN} = g_{MN} + N_M K_N + N_N K_M\, .
\label{TransverseMetric}
\ee
As its name suggests, the transverse metric is also transverse to the null ray $\hat h_{MN} N^M = 0$.
An important feature of the null Raychaudhuri equation (\ref{NullRayEq}) is that it is a purely geometric expression, and is identically true, regardless of the matter content or the form of Einstein's equations. 
For example, it holds for the Einstein equations as well as corrections derived from string theory.
We will find that calculating both sides of the Raychaudhuri equation (\ref{NullRayEq}) for the metric (\ref{exMetric}) imposes strong local constraints on the necessary conditions required to obtain 4-dimensional accelerated expansion.

To begin, an affine null vector for (\ref{exMetric}) along the $t,\chi$ directions is
\be
N^M = \underset{\hspace{.2in}t,\hspace{.1in} \vec{x},\hspace{.1in} \chi,\hspace{.1in} \vec{y}}{\Omega^{-2}\left(1,\ \vec{0},\ 1,\ \vec{0}\right)\, .}
\ee
In order to construct the transverse metric $\hat h_{MN}$ and shear tensor $\sigma_{MN}$ we will need the cross-normalized auxiliary vector
\be
K^M = \frac{1}{2} \left(1,\vec{0},-1,\vec{0}\right)\, .
\ee
The transverse metric (\ref{TransverseMetric}) is then
\be
\hat h_{MN} = \begin{pmatrix}
0 & 0 & 0  & 0\\
0 & \Omega^2 a^2 \delta_{ij} & 0 & 0 \\
0 & 0 & 0 & 0 \\
0 & 0 & 0 & \Omega^2 f^2(\chi) \t g_{mn}
\end{pmatrix}
\ee
and we indeed see that $h_{MN} N^M = 0$.
The expansion of $N^M$ is
\be
\theta = \frac{1}{\sqrt{-g_D}} \partial_A \left(\sqrt{-g_D} N^A\right) = 3H \Omega^{-2} + \Omega^{-2} \partial_\chi \log\left(\Omega^{D-2} f^{n-1}\right)\, .
\label{exTheta}
\ee
From this it is straightforward to compute the derivative with respect to the affine parameter $\lambda$ on the left-hand side of the Raychaudhuri equation (\ref{NullRayEq})
\be
\frac{d\theta}{d\lambda} = N^M \partial_M \theta = N^t \partial_t \theta + N^\chi \partial_\chi\theta = \Omega^{-4} \left[3 \dot H + \partial_\chi^2 \log\left(\Omega^{D-2}f^{n-1}\right)-2 (\partial_\chi \log \Omega)\left(3 H + \partial_\chi \log\left(\Omega^{D-2} f^{n-1}\right)\right)\right]\,.
\label{exThetaLambda}
 \ee
The shear tensor (\ref{ShearTensor}) has non-zero components
\be
\label{exSheartt}
\sigma_{tt} &=& -\partial_\chi \log \Omega\\
\sigma_{t\chi} &=& \sigma_{\chi t} = \partial_\chi \log \Omega \\
\sigma_{tm} &=& \sigma_{mt} = \partial_m \log \Omega \\
\sigma_{ij} &=& \frac{1}{D-2} \left[(D-5) - \partial_\chi \log\left(\Omega^{D-2} f^{n-1}\right)\right] a^2 \delta_{ij} \\
\sigma_{\chi\chi} &=& -\partial_\chi \log \Omega \\
\sigma_{\chi m} &=& \sigma_{m\chi} = \partial_m \log \chi \\
\sigma_{mn} &=& \frac{1}{D-2} \left[-3 H + \partial_\chi \log\left(\Omega^{-2} f^3\right)\right] f^2 \t g_{mn}
\label{exShearmn}
\ee
with all other components vanishing.

Combining (\ref{exTheta}), (\ref{exThetaLambda}), and (\ref{exSheartt})-(\ref{exShearmn}) into the null Raychaudhuri equation (\ref{NullRayEq}), we obtain the remarkably simple expression
\be
3 (\dot H + H^2) = -(D-5) \frac{\partial_\chi^2 f}{f} + (D-2) \left( (\partial_\chi \log \Omega)^2 - \partial_\chi^2 \log \Omega\right) - \Omega^4 R_{MN} N^M N^N\, .
\label{exNoGo1}
\ee
This can be simplified further by noticing that the first term on the right hand side is just the component of the Ricci curvature tensor of the unwarped $\bar{g}_{mn} = f^2(\chi) \t g_{mn}$ metric along the $\chi$ directions, $\bar R_{\chi\chi} = -(D-5) \partial_\chi^2 f/f$, so we can rewrite (\ref{exNoGo1}) as
\be
3 \left(\dot H + H^2\right) = \bar R_{\chi\chi} + (D-2) \left( (\partial_\chi \log \Omega)^2 - \partial_\chi^2 \log \Omega\right) - \Omega^4 R_{MN} N^M N^N\, .
\label{exNoGo2}
\ee
This simple result has important implications: if we wish to have 4-dimensional dS space $\dot H = 0$ or near dS space $|\dot H| \ll  H^2$, then the left hand side of (\ref{exNoGo2}) is positive. Thus, the right-hand side must also be positive. However, this is difficult to achieve because the last term on the right-hand side of (\ref{exNoGo2}) is {\bf always negative} for backgrounds that obey the Null Energy Condition (NEC) $R_{MN} N^M N^N \geq 0$.
We will discuss this in more detail below.

In particular, for a space with non-positive curvature along $\chi$, $\bar R_{\chi\chi} \leq 0$, trivial warping $\Omega = \mbox{const}$, and a background that obeys the Null Energy Condition (NEC), the right-hand side of (\ref{exNoGo2}) is non-positive, so that it is impossible to obtain 4-dimensional dS. It is necessary to violate one of these assumptions in order to get 4-dimensional dS space from a theory with extra dimensions with a metric of the form (\ref{exMetric}).
The result (\ref{exNoGo2}) is quite generic -- we have not made any assumptions about the matter content,
we have not integrated over the extra dimensions or demanded that they are compact or without boundary\footnote{An integrated version of (\ref{exNoGo2}) can be obtained by multiplying both sides by $\Omega^{-4}$ and integrating over the internal dimensions, giving $3(\dot H+H^2) \t V= \int d^ny \sqrt{\t g} \Omega^{-4}[\t R_{\chi\chi}-3(D-2) (\partial_\chi \Omega)^2 - \Omega^4 R_{MN} N^M N^N]$ where $\t V = \int d^ny \sqrt{\t g} \Omega^{-4}$. The warp factor term now contributes non-positively to the right-hand side, so one must have some combination of positive curvature and NEC violation in order to evade this integrated version of (\ref{exNoGo2}).}, and we have not even assumed the Einstein equations. The result (\ref{exNoGo2}) merely follows directly from the geometric identity of the Raychaudhuri equation (\ref{NullRayEq}).

The simple form of (\ref{exNoGo2}) can hide some of its important features.
Since (\ref{exNoGo2}) is not integrated over the internal space, it applies equally well to compact as well as non-compact extra dimensions, as well as extra dimensions that are terminated on boundaries such as branes.
Not being integrated, the left-hand side of (\ref{exNoGo2}) is independent of the extra dimensions, so the right-hand side of (\ref{exNoGo2}) must also be independent of the extra dimensions as well.
This makes it difficult to imagine how the second term on the right-hand side of (\ref{exNoGo2}), involving the warp factors, can play an important role, except for very special solutions.
Indeed, we will find in specific examples that this warp factor term is either constant (for specific solutions), exactly zero, or cancels against corresponding terms arising from the NEC term so that it does not provide a net positive contribution to the right-hand side.
Finally, (\ref{exNoGo2}) must apply locally at every point in the extra dimensions.
This implies that localized violations of the NEC, such as through local sources (like branes) with support only on a submanifold $\Sigma$ of the internal space, are insufficient for satisfying (\ref{exNoGo2}) homogeneously at every point in the extra dimensions.
While it is possible that a NEC-violating source with only local support on $\Sigma$ may give a positive contribution to the integral of $-R_{MN} N^M N^N$ over the internal space, this would be insufficient to satisfy (\ref{exNoGo2}) at every point in the extra dimensions, since the contribution from this source would vanish for points not located on the submanifold $\Sigma$.

Let us now consider a more general warped product metric
\be
ds^2 &=& \Omega^2(y^m) \left[\hat g_{\mu\nu} dx^\mu dx^\nu + \t g_{mn}(y) dy^m dy^n\right] \nonumber \\
&=& \Omega^2(y^m) \left[-dt^2 + a^2(t) \delta_{ij} dx^i dx^j + \t g_{mn}(y) dy^m dy^n\right]\, ,
\label{GeneralMetric}
\ee
with affine null vector with legs in the extra dimensions,
\be
N^M =\Omega^{-2} \underset{t,\hspace{.1in} \vec{x}, \hspace{.1in} \vec{y}}{\left(1,\ \vec{0},\ \t n^m\right)}\, ,
\ee
where $\t n^m$ is a unit vector with respect to $\t g_{mn}$, $\t n^m \t n^n \t g_{mn} = 1$, and is affine $\t n^m \t \nabla_m \t n^n = 0$.
Our previous metric (\ref{exMetric}) can be seen as a particular special case of (\ref{GeneralMetric}), in which we pull out the radial direction $\chi$.
The Raychaudhuri equation (\ref{NullRayEq}) for this metric (\ref{GeneralMetric}) becomes
\be
3\left(\dot H + H^2\right) = \t R_{mn} \t n^m \t n^n + (D-2) \t n^m \t n^n \left((\partial_m \log \Omega)(\partial_n \log \Omega) - \t \nabla_m \partial_n \log \Omega\right) - \Omega^4 R_{MN} N^M N^N\, .
\label{GeneralNoGo}
\ee
This more general constraint equation (\ref{GeneralNoGo}) reduces to (\ref{exNoGo2}) when we take $\t g_{\chi\chi} = 1$ and take the extra-dimensional null vector component to be along the $\chi$-direction.
Note that (\ref{GeneralNoGo}) shares the local, un-integrated properties of (\ref{exNoGo2}) and shares the same constraint condition: for non-positive curvature of the extra dimensions, constant warping, and non-violation of the NEC, it is not possible to have 4-dimensional de Sitter space with the generic metric (\ref{GeneralMetric}).

We stress again that the constraints
(\ref{exNoGo2}),(\ref{GeneralNoGo}) are purely based on the geometry of their corresponding metrics (\ref{exMetric}),(\ref{GeneralMetric}), and do not make any assumptions about the matter content nor use the Einstein equations.
However, since the last term involving the NEC plays an important role, it is helpful to recast this in terms of a constraint on the matter content of the theory through the Einstein equations.
Using the Einstein equations, we can write the NEC in its more familiar form as a condition on the matter content of the theory, $R_{MN} N^M N^N = \kappa_D^2 T_{MN} N^M N^N \geq 0$. 
The NEC is obeyed by essentially all classical forms of matter, making it difficult to realize 4-dimensional dS space in (\ref{exNoGo2}),(\ref{GeneralNoGo}) using known matter ingredients\footnote{It may be possible to violate the NEC with exotic forms of matter \cite{ArkaniHamed:2003uz,Kobayashi:2010cm}, non-minimal coupling \cite{Lee:2007dh,Lee:2010yd}, or quantum gravity effects \cite{Ford:1993bw}, though these approaches often face challenges that we will not explore further here.}.
For example, a bulk $D$-dimensional cosmological constant with energy momentum tensor
\be
T_{MN}^\Lambda = - \Lambda_D\  g_{MN}
\ee
saturates the NEC, $T_{MN}^\Lambda N^M N^N = 0$, and therefore does not contribute at all to (\ref{exNoGo2}),({\ref{GeneralNoGo}).
A canonically normalized bulk scalar field $\phi$ with potential $V(\phi)$ has energy-momentum tensor
\be
T_{MN}^\phi = 2\partial_M \phi \partial_N \phi - g_{MN} \left((\partial\phi)^2 + V(\phi)\right)
\ee
which is non-negative when contracted with the null vector $T_{MN}^\phi N^M N^N = \left(N^M\partial_M \phi \right)^2 \geq 0$. Thus it contributes as a negative term to the right-hand side of (\ref{exNoGo2}),(\ref{GeneralNoGo}).
Fluxes arising from the $p$-form field strength tensor $F_{a_1...a_p}$ have the energy-momentum tensor
\be
T_{MN}^p = 2p\ F_{Ma_2...a_p}F_N^{a_2...a_p} - g_{MN} F^2\, .
\ee
If the $p$-form flux is purely along the extra directions, and not along the direction of the null vector, then we find that the fluxes saturate, but do not violate, the NEC. If the flux is along the extra dimensions and has components along the direction of the null vector, we see that the flux contribution is non-negative and does not violate the NEC.
Finally, if the $p$-form flux is spacetime-filling in the 4-dimensions (and for $p>4$ has additional components in the extra dimensions), we again see that the flux contribution does not violate the NEC.
Thus, $p$-form fluxes cannot provide a positive contribution to the right-hand side of (\ref{exNoGo2}),(\ref{GeneralNoGo}).

A common source of NEC-violation in string flux compactifications is that of $(p+1)$-dimensional local sources with tension $T_p$ and energy momentum tensor
\be
T_{\mu\nu}^{loc} = - T_p g_{\mu\nu} \delta(\Sigma), \hspace{.1in} T_{mn}^{loc} = - T_p\, \Pi_{mn}^\Sigma \delta(\Sigma), \hspace{.1in} \Pi_{mn}^\Sigma g^{mn} = p-3 
\ee
which are localized on the submanifold $\delta(\Sigma)$, and depends on the projector $\Pi_{mn}^\Sigma$ onto this cycle $\Sigma$.
For points in the extra dimensions away from $\Sigma$, the energy-momentum tensor of these localized sources vanishes.
If the tension of such objects is negative $T_p = -|T_p| < 0$ (such as for orientifold-planes), the {\bf integrated} version of the null energy condition
\be
\int d^n y\ T_{MN}^{loc} N^M N^N \sim T_p < 0
\ee
can be violated.
However, as discussed above, the constraints (\ref{exNoGo2}),(\ref{GeneralNoGo}) are local, so that that the constraints must be satisfied point-by-point throughout the extra dimensions; in particular, they must be satisfied away from the submanifold $\Sigma$ where localized sources do not have support, and whose local contributions vanish.
Thus, localized sources, even those that violate the NEC, will not contribute directly to satisfying the constraint.
A common practice when constructing string flux compactifications is to {\it smear} the localized sources, dissolving their tension and charge uniformly throughout the extra dimensions. While this may be useful for technical considerations, it side-steps one of the central challenges of our constraints to these models, in that the constraints must be satisfied point-by-point throughout the extra dimensions so that the NEC must be violated at every point.
While it remains an interesting open question whether smeared solutions survive intact upon localization, the requirement that our constraints must be satisfied point-by-point presents a potential obstacle to realizing dS solutions upon localization.

\subsection{Relation to Other No-Go's}

Our constraints (\ref{exNoGo2}),(\ref{GeneralNoGo}) bear some resemblance to other existing ``no-go'' theorems in the literature for obtaining dS space in models with warped extra dimensions.
Before we continue to examine the implications of the constraints, then, we will remark briefly on the comparison of (\ref{exNoGo2}),(\ref{GeneralNoGo}) to these previous results.

One set of ``no-go'' theorems for dS space\footnote{There are also similar ``no-go'' theorems for time-dependent compactifications \cite{Townsend:2003fx,Teo:2004hq,Russo:2018akp,Russo:2019fnk}, though we will focus on the time-independent case here.} compactifications arises by taking the 4-dimensional trace of the trace-reversed Einstein equations from the metric (\ref{GeneralMetric}), leading to \cite{Gibbons:2003gb,MN,Douglas:2010rt,Dasgupta:2014pma}
\be
\hat R_4 &=& \t \nabla^2 \log \Omega + (D-2) \left(\t \nabla \log\Omega\right)^2 - \Omega^2 \t T \nonumber \\
&=& \frac{1}{(D-2) \Omega^{D-2}} \t \nabla^2 \Omega^{D-2} - \Omega^2 \t T
\label{MNNoGo}
\ee
where
\be
\t T = \frac{1}{D-2} \left[-(D-6) T^\mu_\mu + 4 T^m_m\right]\, .
\label{MNT}
\ee
(In more recent literature \cite{Hamada:2019ack,Carta:2019rhx}, this combination is commonly written as $\Delta$.)
There are clear similarities between (\ref{MNNoGo}) and (\ref{GeneralNoGo}): in order to have 4-dimensional dS space, the left-hand side of (\ref{MNNoGo}) is positive so that the right-hand side must be positive as well, suggesting that $\t T < 0$ is a necessary condition for obtaining dS space. Notice, however, that (\ref{MNNoGo}) does not (directly) depend on the curvature of the internal space, in contrast to (\ref{GeneralNoGo}).
Interestingly, previous studies have used (\ref{MNNoGo}) to infer
that {\it negative} internal curvature is a necessary condition for obtaining dS space \cite{Douglas:2010rt}.
We can see clearly from (\ref{GeneralNoGo}), however, that negative curvature will not help evade this constraint, but instead will make it worse. Instead, positive internal curvature arises as a sufficient (but perhaps not necessary) condition for dS space.

Multiplying (\ref{MNNoGo}) by $\Omega^{D-2}$ and integrating over the internal space gives
\be
\mbox{\bf GMN No-Go} \hspace{.4in} \t {\mathcal V} \hat R_4 = \frac{1}{D-2} \int d^{D-4} y \sqrt{\t g}\, \t \nabla^2 \Omega^{D-2}- \int d^{D-4} y \sqrt{\t g}\, \Omega^D\,  \t T\, .
\label{MNIntegrated}
\ee
It is tempting to set the first term on the right-hand side of (\ref{MNIntegrated}) to zero, since it appears to be a total derivative on a compact space.
However, because of potential singularities in the warp factor arising from branes or orientifold planes, this term could be non-zero (see \cite{MN,Dasgupta:2014pma} for more discussion). We will refer to the integrated no-go constraint (\ref{MNIntegrated}), based on the analysis of \cite{Gibbons:2003gb,MN}, as the Gibbons-Maldacena-Nunez (GMN) No-Go.

In order for the GMN No-Go (\ref{MNIntegrated}) to give rise to 4-dimensional dS, the Energy-Momentum tensor must satisfy
\be
\t T < 0 \hspace{.2in} \Rightarrow \hspace{.2in} (D-6) T^\mu_\mu - 4 T^m_m > 0\, ,
\label{MNTCondition}
\ee
which we will compare to constraints arising from (\ref{GeneralNoGo}).
It is readily seen that fluxes do not satisfy (\ref{MNTCondition}) \cite{MN}, while positive tension branes also fail for $D>6$ \cite{Dasgupta:2014pma}. 

As an example, consider models with negative internal curvature, ordinary fluxes, no warping, and localized objects such as branes and orientifolds, which have been proposed to lead to 4-dimensional dS.
Negative curvature has been thought to be an important component in a successful construction of dS with extra dimensions \cite{Saltman:2004jh,Silverstein:2007ac,Douglas:2010rt}.
Indeed, these ingredients appear to satisfy the condition (\ref{MNTCondition}) (or its integrated version (\ref{MNIntegrated})), implying that 4-dimensional dS solutions are potentially possible.
Our constraint (\ref{GeneralNoGo}), however, shows that these putative solutions are not solutions with 4-dimensional dS after all, since they do not have positive curvature, warping, or matter content which violates the NEC throughout the extra dimensions.
More generally, it appears that the presence of negative curvature makes it more difficult to obtain a dS solution, since it contributes negatively to the right-hand side of (\ref{GeneralNoGo}), and that positive curvature instead appears to be preferred.
We will illustrate this further with an explicit solution of 4-dimensional dS space with positive internal curvature in the following subsection.

As more evidence that our constraint (\ref{GeneralNoGo}) excludes ingredients that may have otherwise passed previous no-go's, notice that
a $D$-dimensional cosmological constant $T_{MN}^\Lambda = -\Lambda_D\, g_{MN}$ satisfies (\ref{MNTCondition})
\be
\t T^\Lambda = -(D-6) \left(T^\Lambda\right)^\mu_\mu + 4 \left(T^\Lambda\right)^m_m = -8 \Lambda_D < 0
\ee
as long as $\Lambda_D > 0$, suggesting that it could be possible to obtain 4-dimensional dS with a $D$-dimensional cosmological constant.
However, as we have seen, a $D$-dimensional cosmological constant saturates, but does not violate, the NEC, so that a $D$-dimensional cosmological constant is not sufficient in (\ref{GeneralNoGo}) to allow for 4-dimensional dS.
Thus, the constraints arising from (\ref{MNTCondition}),(\ref{MNIntegrated}) are weaker than those arising from (\ref{GeneralNoGo}), which requires more than a $D$-dimensional cosmological constant in order to satisfy the conditions for dS space.
The constraints (\ref{MNIntegrated}) are also weaker because they are integrated over the internal space, while (\ref{GeneralNoGo}) are valid at every point throughout the internal space.
Even working with the un-integrated version (\ref{MNNoGo}), however, requires the assumption of the Einstein equations, which are not assumed in the derivation of (\ref{GeneralNoGo}).

Our results (\ref{exNoGo2}),(\ref{GeneralNoGo}) also bear some resemblance to work by Steinhardt-Wesley \cite{Wesley:2008fg,Steinhardt:2008nk} (see also \cite{Obied:2018sgi}), in which the NEC also plays an important role.
While we are interested primarily in 4-dimensional dS space, \cite{Wesley:2008fg,Steinhardt:2008nk} are interested more generally in accelerating cosmologies with a more general time-dependence, and specifically require integration over the internal dimensions in order to draw conclusions.
A list of assumptions necessary to derive the results of \cite{Steinhardt:2008nk} are conveniently summarized in \cite{Postma}; in Table \ref{tab:Compare}, we briefly review some of the assumptions from \cite{Gibbons:2003gb,MN} and \cite{Steinhardt:2008nk} and how they compare to our assumptions.
We see in general that our constraint requires fewer assumptions than previous no-gos.

\begin{table}[t]
\setlength{\tabcolsep}{0.5em}
\renewcommand{\arraystretch}{1.5}
\begin{tabular}{p{3in}|>{\centering\arraybackslash}p{.5in}|>{\centering\arraybackslash}p{.8in}|c}
{\bf Assumption} & {\bf GMN No-Go} & {\bf Steinhardt-Wesley} & {\bf This Work} \\ \hline
The metric is block-diagonal & Yes & Yes & Yes \\ \hline
Higher dimensional theory is assumed to be described by General Relativity & Yes & Yes & No \\ \hline
The internal metric is Ricci-flat or conformal Ricci-flat & No & Yes & No \\ \hline
The integral of a total derivative over the internal space vanishes or is bounded & Yes & Yes & No \\ \hline
\end{tabular}
\caption{A comparison of some of the main assumptions of this work compared to assumptions of other well-known no-go theorems for obtaining accelerating 4-dimensional cosmologies with extra dimensions.}
\label{tab:Compare}
\end{table}

\subsection{Example: Freund-Rubin with Cosmological Constant}
\label{sec:FR}

The constraint (\ref{exNoGo2}) appears to be rather stringent, since as we have seen there are very few energy-momentum sources which violate the NEC in a uniform way throughout the extra dimensions.
However if the internal space has {\it positive curvature} then it is possible, from the perspective of (\ref{exNoGo2}), to obtain a compactification to dS$_4$.

In this section we will show that this expectation is indeed true by presenting a simple example, following \cite{FreundRubin,RandjbarDaemi:1982hi,Carroll:2001ih}, consisting of 2-form flux and a bulk cosmological constant on a 2-sphere.
We will examine this model from the perspective of the 4-dimensional effective action, the 6-dimensional Einstein equations, and the Raychaudhuri constraint (\ref{exNoGo2}) in turn, demonstrating agreement between each perspective.

We will take an unwarped direct product metric for a $D=6$ compactification to $4$-dimensions on a 2-sphere $S^2$:
\be
ds_6^2 = L^{-2}\hat g_{\mu\nu} dx^\mu dx^\nu + L^2\t g_{mn} dy^m dy^n
\label{FRMetric}
\ee
where 
$L$ is the radius of the $S^2$, and we have done a Weyl rescaling so that the 4-dimensional Planck constant is independent of $L$.
Taking the matter content to be a 2-form gauge field strength $F_2$ wrapped over the extra dimensions, $F_2 = f_2\, \t \epsilon_2$ (where $\t \epsilon_2$ is the volume form on $\t g_{mn}$), and a 6-dimensional cosmological constant $\Lambda_6$, the 6-dimensional action is
\be
S_6 = \int d^6x \sqrt{-g_6} \left[M_6^4 R_6 - \Lambda_6 - \frac{1}{4} F_2^2\right]\, .
\label{FRAction}
\ee
An effective potential for the radius $L$ can be found by dimensionally reducing this action (\ref{FRAction}),
\be
S_4 = \int d^4x \sqrt{-\hat g_4} \left[M_p^2 \hat R_4 - V_{eff}(L)\right]
\ee
where $M_p^2 = \t V_2\, M_6^4$ and $\t V_2 = \int d^2y \sqrt{\t g_2}$, with
\be
V_{eff}(L) = -M_p^2\frac{\t R_2}{L^4} + \frac{\t V_2}{2} \frac{f_2^2}{L^6} + \t V_2 \frac{\Lambda_6}{L^2}
\label{FREffPot}
\ee
where $\t R_2 = \t g^{mn}\t R[\t g]_{mn}$ is the fiducial curvature of the $S^2$. 
The minimum of the potential is found at
\be
\partial_L V_{eff}(L) = 0 \hspace{.2in}\Rightarrow  \hspace{.2in} 4 \frac{M_p^2 \t R_2}{L^5} - 3\frac{\t V_2 f_2^2}{L^7}-2\frac{\t V_2 \Lambda_6}{L^3} = 0\, .
\label{FREffPotMin}
\ee

\begin{figure}[t!]
\centering\includegraphics[width=.6\textwidth]{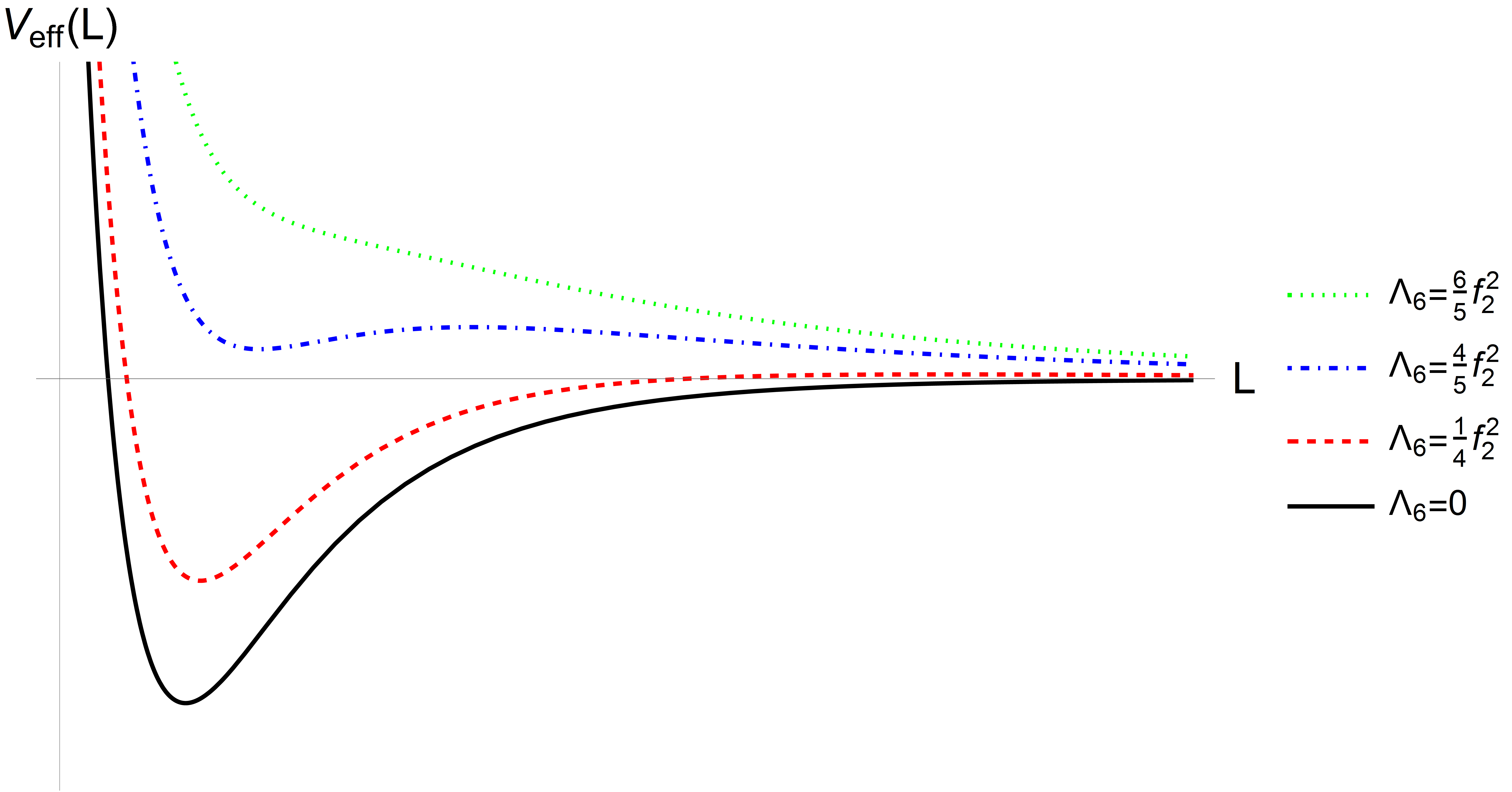}
\caption{The effective potential (\ref{FREffPot}) for various values of the bulk cosmological constant $\Lambda_6$ in units of $M_6 = 1$. Notice that the cosmological constant not only ``uplifts'' the effective potential, but also shifts the location of the stable minimum.}
\label{fig:FREffectivePotential}
\end{figure}

In the absence of a bulk cosmological constant $\Lambda_6 = 0$, the effective potential (\ref{FREffPot}) has a stable minimum at finite radius with a negative vacuum energy, as in Figure \ref{fig:FREffectivePotential}; this is the usual Freund-Rubin solution.
The presence of the bulk cosmological constant has two important effects, which can be readily seen in Figure \ref{fig:FREffectivePotential}: First, it appears to contribute a positive ``uplifting'' energy density, and can lift the negative vacuum energy at the stable minimum to a positive vacuum energy with a (meta)-stable minimum;
second, the functional dependence of the cosmological constant contribution slightly shifts the value of the stabilized radius away from the negative cosmological constant minimum to larger values. If the bulk cosmological constant is too large, however, the meta-stable de Sitter minimum disappears.
We can see this second point directly from the minimization of the potential (\ref{FREffPotMin}):
increasing $\Lambda_6$ from zero shifts the minimum to larger values of $L$, until the critical value $\Lambda_6 \geq \frac{2}{3} M_6^8 \frac{\t R_2^2}{f_2^2}$ is reached and a minimum no longer exists, corresponding to the vanishing of the uplifting term.

Now let's consider the same system, but from the perspective of the Einstein equations.
From (\ref{FRAction}), the 6-dimensional Einstein equations reduce to the two conditions
\be
\label{FR_Einstein1}
\hat R_4 &=& -2 \frac{\t R_2}{L^4} + \frac{1}{M_6^4} \frac{f_2^2}{L^6} + \frac{2}{M_6^4} \frac{\Lambda_6}{L^2} \\
\t R_2 &=& \frac{3}{4M_6^4} \frac{f_2^2}{L^2} + \frac{1}{2 M_6^4} \Lambda_6\ L^2
\label{FR_Einstein2}
\ee
in terms of the 4-dimensional curvature $\hat R_4 = \hat g^{\mu\nu}\hat R[\hat g]_{\mu\nu}$.
Notice that, up to multiplicative constants, the first of these (\ref{FR_Einstein1}) corresponds to the effective potential (\ref{FREffPot}), while the second (\ref{FR_Einstein2}) corresponds to the minimization condition (\ref{FREffPotMin}).
Thus, we see that the 6-dimensional Einstein equations are completely equivalent to the 4-dimensional effective potential perspective.

Now, let's consider this simple model in the context of the constraint (\ref{exNoGo2}). Specifically, we will take the 2-sphere metric to be
\be
\t g_{mn} dy^m dy^n = d\chi^2 + \sin^2\chi d\phi^2\, ; 
\ee
this metric has the same form as (\ref{exMetric}), and we will take our null ray to have its spatial leg along the $\chi$ direction, so that the constraint (\ref{exNoGo2}) can be used with no change in its form.
Evaluating the right-hand side of (\ref{exNoGo2}), 
\be
3 H^2 = \frac{1}{L^4} \t R_{\chi\chi} - R_{MN} N^M N^N = \frac{1}{2} \frac{\t R_2}{L^4} - \frac{1}{2 M_6^4} \frac{f_2^2}{L^6}\, ,
\label{FRNoGo1}
\ee
where we used $R_{MN} N^M N^N = \frac{1}{M_6^4}T_{MN} N^M N^N$,
we find that the positive curvature contributes a positive term to the right-hand side, the warp factor terms vanish, and the $F_2$ flux contributes a negative term through the energy-momentum tensor. Notice that the contribution from the cosmological constant vanishes since it saturates the NEC.

Writing the constraint in this form presents a puzzle: how does the introduction of the bulk cosmological constant $\Lambda_6$ lead to a positive $H^2 > 0$ if it doesn't contribute to the right-hand side of (\ref{FRNoGo1})?
The solution to this puzzle is that the bulk cosmological constant indirectly affects the balance of terms on the right-hand side of (\ref{FRNoGo1}). When $\Lambda_6 = 0$, the location of the stable minimum is such that the second term on the right-hand side of (\ref{FRNoGo1}) is larger than the first, so that the right-hand side is overall negative. However, as $\Lambda_6$ is increased from zero, the location of the stable minimum shifts to larger $L$. Since the first term is proportional to $L^{-4}$ while the second term is proportional to $L^{-6}$, as $L$ increases the magnitude of the first term overtakes the second, so that at large enough values of $\Lambda_6$ the stable minimum is located at a large enough radius for $L$ that (\ref{FRNoGo1}) becomes positive.
This ``indirect uplifting'' challenges the usual perspective of simply adding a positive contribution to the effective potential to uplift the minimum to a de Sitter minimum. Instead, the additional ingredient of a cosmological constant shifts the location of the minimum in the right direction to result in a de Sitter solution that satisfies (\ref{exNoGo2}).

We have evaded the constraint (\ref{exNoGo2}), obtaining a stabilized compactification with 4-dimensional de Sitter space, through the simple presence of positive curvature, 2-form gauge field strength, and a bulk cosmological constant.
While it is possible this model may not be directly realizable as the low-energy limit of a top-down construction (such as string theory), it is nonetheless useful as a concrete, explicit example of a model that evades the constraint (\ref{exNoGo2}) as well as other no-go theorems for dS with extra dimensions.

In particular, we have shown explicitly that positive (not negative) curvature is the necessary ingredient in this model to obtain 4-dimensional dS.
While it is tempting to think of the bulk cosmological constant as the essential ingredient that uplifts the minimum to de Sitter, we found instead through the constraint that since the bulk cosmological constant saturates the NEC it does not directly contribute to the vacuum energy at the minimum (\ref{FRNoGo1}).
Instead, (\ref{FRNoGo1}) illustrates that the bulk cosmological constant contributes indirectly: by shifting the location of the stable minimum to larger values of the radius the positive curvature overtakes the negative contribution from the 2-form flux in (\ref{FRNoGo1}).

\subsection{Example: dS in RS}
\label{sec:dSinRS}

Our simple example in Section \ref{sec:FR} had a constant warp factor, so that the contributions from warping to (\ref{exNoGo2}),(\ref{GeneralNoGo}) vanished.
More generally, however, we expect some non-trivial warping due to e.g.~localized sources or fluxes in the extra dimensions. While the details of the warping depends in many cases on the specific ingredients, the most well-known and explicit constructions of extra dimensions with warping are the Randall-Sundrum (RS) models \cite{Randall:1999ee,RS2}, with the warped metric:
\be
ds^2_{RS} = e^{2A(z)} \hat g_{\mu\nu} dx^\mu dx^\nu + dz^2\, .
\label{RSMetric1}
\ee
The 5-dimensional theory is described by $5$-dimensional General Relativity with a bulk cosmological constant
\be
\int d^5x \sqrt{-g_5}\left[\frac{M_5^2}{2} R_5 - \Lambda_5\right]
\ee
where the extra dimension along $z$ is an $S^1$ terminated by the presence of one or more 3-branes with brane tensions $\lambda_i$ for $i=1,2$.
The presence of the boundaries makes it impossible to apply the integrated no-go theorems \cite{MN,Douglas:2010rt,Dasgupta:2014pma,Wesley:2008fg,Steinhardt:2008nk}; however, as previously discussed the constraint (\ref{exNoGo2}) applies equally well in the presence of boundaries and is true everywhere in the bulk locally, so it can be used to analyze RS backgrounds.

When the brane tensions are tuned to be related to the (negative) bulk cosmological constant
$\Lambda_5 = -6 M_5^2/L^2$,
$\lambda_1 = -\lambda_2 = 6 M_5^3/L$,
the warp factor takes the form $A(r) = -z/L$ and the 4-dimensional metric is Minkowsi \cite{Randall:1999ee}.
A simple (and, as we will see, overly naive) choice for this metric is to take the RS solution $A(z) = -z/L$ and to promote the 4-dimensional metric $\hat g_{\mu\nu}$ to be expanding
\be
ds^2_{RS} = e^{-2z/L} \left[-dt^2 + a^2(t) d\vec{x}^2\right] + dz^2\, .
\label{RSMetric2}
\ee
We can rewrite (\ref{RSMetric2}) in the form (\ref{exMetric})
\be
ds_{RS}^2 = \frac{L^2}{\chi^2} \left[-dt^2 + a^2(t) d\vec{x}^2 + d\chi^2\right]
\label{RSMetric3}
\ee
with $\chi = \frac{1}{L} e^{z/L}$ and $\Omega(\chi) = L/\chi$.
Taking a null vector with one leg along the bulk direction $N^M = \chi^2 \left(1,\vec{0},1\right)$,
the warp factor contribution to (\ref{exNoGo2}) is
\be
\left(\partial_\chi \log \Omega\right)^2 - \partial_\chi^2 \log \Omega = 0
\label{RSWarpContribution}
\ee
so that the constraint equation (\ref{exNoGo2}) becomes (with $\dot H = 0$)
\be
3 H^2 = -\frac{1}{\chi^4} R_{MN} N^M N^N\, .
\label{RSNoGo}
\ee

We see that 4-dimensional dS in RS, with a simple warped product metric of the form (\ref{RSMetric2}), requires a violation of the NEC everywhere in the bulk.
The RSI scenario \cite{Randall:1999ee} requires a 3-brane with negative tension localized at $z = z_0$, which in principle violates the NEC at that location; however, as mentioned above, the constraint (\ref{RSNoGo}) is a local expression, valid at each point in the bulk, even away from the 3-branes, so these do not contribute except as delta-function sources.
RS models also typically involve a number of other ingredients in the bulk, such as bulk scalar fields contributing to the Goldberger-Wise stabilization of the radion \cite{GoldbergerWise}. However, we have already seen that bulk scalar fields do not violate the NEC, and so cannot lead to a positive $H^2$ in (\ref{RSNoGo}).

The analysis leading up to (\ref{RSNoGo}) was done assuming that the warp factor took its Minkowski-space form (\ref{RSMetric2}), and did not change when ``uplifted'' to dS.
However, braneworld and RS models with realistic cosmologies require modifications to the metric that go beyond simply promoting the 4-dimensional metric to dS.
For example, models with dS or AdS space require modifications to the warp factor away from that of flat space \cite{Kaloper:1999sm,DeWolfe:1999cp,Kim:1999ja,Tye:2000fw,Karch:2000ct}, while braneworld cosmological metrics contain warp factors with non-separable dependence on time and the extra dimension \cite{Binetruy:1999hy,Maartens:2010ar}.
We can see the difficulty in obtaining 4-dimensional dS space for the metric (\ref{RSMetric2}) without violating the NEC as a failure of the approach of simply promoting the 4-dimensional metric to be expanding, without also modifying the warp factor.

In particular, by ``de-tuning'' the brane tensions from the Minkowski solution above,
we can obtain solutions with dS or AdS worldvolumes \cite{Kaloper:1999sm,DeWolfe:1999cp,Kim:1999ja,Tye:2000fw,Karch:2000ct}.
Following \cite{DeWolfe:1999cp}, we will take a 5-dimensional spacetime with a bulk cosmological constant $\Lambda_5 = -6 M_5^3/L^2$, bounded by a positive tension brane, with tension $\lambda_1$, located at $z=0$ and negative tension brane, with tension $\lambda_2$, located at $z_0$. 
Unlike the flat RS case, however, now the brane tensions are not tuned to cancel against the bulk cosmological constant.
With the metric (\ref{RSMetric1}), $\hat g_{\mu\nu}$ is 4-dimensional dS with effective 4-dimensional cosmological constant $\Lambda_4$ when
the warp factor and brane tensions take the form
\be
e^{A(z)} &=& \sqrt{\Lambda_4} L\ \sinh\left(\frac{z_\star-z}{L}\right)\, , \hspace{.3in} \lambda_1 = \frac{6 M_5^3}{L} \coth\frac{z_\star}{L}\, ,\hspace{.3in} \lambda_2 = -\frac{6 M_5^3}{L} \coth\frac{z_\star-z_0}{L}\, ,
\label{dWWarpFactor}
\ee
where $z_\star$ is a constant determined in terms of the brane tensions.
For this solution the bulk matter only consists of a 5-dimensional cosmological constant, so $R_{MN} N^M N^N = \kappa_5^2 T_{MN}^\Lambda N^M N^N = 0$. 
The Ricci curvature vanishes, so the constraint (\ref{exNoGo2}) becomes
\be
3 H^2 = 3 \left[\left(\partial_\chi \log \Omega\right)^2 - \partial_\chi^2 \log\Omega\right]\, .
\label{RSdWNoGo}
\ee
Making the coordinate change $dz/d\chi = e^{A(z)}$ with $\Omega = e^{A(z)}$, we find
\be
\partial_\chi \log \Omega &=& -\sqrt{\Lambda_4} \cosh\left(\frac{z_1-z}{L}\right) \\
\partial_\chi^2 \log \Omega &=& \Lambda_4 \sinh^2 \left(\frac{z_1-z}{L}\right)\, ,
\ee
so that (\ref{RSdWNoGo}) becomes
\be
H^2 = \Lambda_4 \cosh^2\left(\frac{z_1-z}{L}\right) - \Lambda_4 \sinh^2\left(\frac{z_1-z}{L}\right) = \Lambda_4\, .
\ee
Thus, we see in this case that the warp factor plays a crucial role: while the internal curvature vanishes and the bulk cosmological constant saturates the NEC, it is the warp factor\footnote{As noted in \cite{Kaloper:1999sm,DeWolfe:1999cp}, we can also interpret these dS solutions as embeddings of dS$_4$ hypersurfaces in AdS$_5$.} itself that sources 4-dimensional dS space.
It is also possible to construct more general solutions with bulk scalar fields \cite{DeWolfe:1999cp}, with a similar role played by the warp factor in evading the constraint (\ref{exNoGo2}).

As in the previous section, while the toy-model example presented here may not be 
directly realizable in a top-down construction from string theory or UV-complete theory of gravity, it nevertheless serves as a proof of principle and guidepost for evading the constraint (\ref{GeneralNoGo}) through the warp factor.
This example also suggests a more general lesson for model building of cosmologies with warped extra dimensions: it may not be enough to simply promote a 4-dimensional metric to dS, leaving the warp factor and other fields fixed. 
Instead, the warp factor may itself play an important role in establishing the existence of 4-dimensional dS.

\subsection{Discussion}

\subsubsection*{Flux Compactifications \& KKLT}

A broad class of warped metrics used in string compactifications take the form \cite{GKP}
\be
ds_{10}^2 = e^{2A(y^m)} \hat g_{\mu\nu} dx^\mu dx^\nu + e^{-2A(y^m)} \bar g_{mn} dy^m dy^n\, .
\label{GKPMetric1}
\ee
We can rewrite the constraint (\ref{GeneralNoGo}) in terms of the warp factor $A(y)$ and ``unwarped'' metric $\bar g_{mn}$,
\be
3\left(\dot H + H^2\right) = \bar R_{mn}\t n^m \t n^n + 2 e^{4A} \bar \nabla^2 A - 8 \t n^m \t n^n \partial_n A \partial_m A - e^{4A} R_{MN} N^M N^N
\label{GKPNoGo1}
\ee
where $\bar R_{mn}$ is the Ricci tensor of the metric $\bar g_{mn}$, we used $\Omega = e^A$, and internal derivatives are with respect to the unwarped coordinates.
In order to determine the precise form of the contribution of the warp factor to (\ref{GKPNoGo1}), we need to consider a specific model.

A common form of (\ref{GKPMetric1}) with warping of the form 
\be
ds_{10}^2 = \frac{r^2}{L^2} \hat g_{\mu\nu} dx^\mu dx^\nu + \frac{L^2}{r^2} \left(dr^2 + r^2 dS_5^2\right)
\label{AdS5Metric1}
\ee
serves as a simple stand-in for many other more complicated strongly warped backgrounds, such as the Klebanov-Strassler throat \cite{Klebanov:2000hb}.
It is straightforward to factor out a conformal warp factor from the metric (\ref{AdS5Metric1})
\be
ds_{10}^2 = \frac{L^2}{\chi^2} \left(-dt^2 + a^2(t) d\vec{x}^2 + d\chi^2 + \chi^2 dS_5^2\right)
\label{AdS5Metric2}
\ee
where $\chi = L/r$, so that (\ref{AdS5Metric2}) takes the form (\ref{exMetric}) with $\Omega(\chi) = L/\chi$.
We then find that the warp factor contribution to the constraint (\ref{exNoGo2}) vanishes
\be
\left(\partial_\chi \log \Omega\right)^2 - \partial_\chi^2 \log\Omega = \left(\partial_\chi \left(-\log \chi\right)\right)^2 - \partial_\chi^2 \left(-\log \chi\right) = 0
\ee
so that (\ref{exNoGo2}) becomes:
\be
3 H^2 = -\frac{L^4}{\chi^4} R_{MN} N^M N^N\, .
\ee
As we saw in Section \ref{sec:dSinRS}, simply promoting the 4-dimensional metric of a known warped solution to dS while leaving the warp factor the same requires that the NEC is violated at every point throughout the bulk of the extra dimensions.

Another well-studied model based on (\ref{GKPMetric1}) that purports to lead to 4-dimensional dS space is KKLT \cite{KKLT}.
The KKLT construction consists of three separate sets of ingredients: First, a low energy ${\mathcal N}=1$ 4-dimensional effective field theory is calculated based on the metric (\ref{GKPMetric1}) with a Ricci-flat internal space within type IIB string theory with fluxes and local sources such as D3-branes and O3-planes, as in \cite{GKP}.
The fluxes and branes source the warp factor and stabilize the complex structure moduli, fixing the dynamics of the internal space.
Second, gaugino condensation on D7-branes give rise to a non-perturbative contribution to the superpotential of the 4-dimensional effective field theory, stabilizing the overall volume modulus of the internal space.
Finally, an anti D3-brane located in a strongly warped region of the internal space uplifts the minimum of the effective potential to a positive cosmological constant.

Let's examine this model and its ingredients in the context of our constraint (\ref{GKPNoGo1}).
As we have argued above, we see from (\ref{GKPNoGo1}) that any model with non-positive curvature, trivial warping, and matter sources that obey the NEC cannot give rise to 4-dimensional dS.
For the first step of the construction, we have already seen that fluxes do not violate the NEC, and so cannot contribute positively to the right-hand side of (\ref{GKPNoGo1}), and since the internal space is Ricci-flat $\bar R_{mn} = 0$ the first term on the right-hand side of (\ref{GKPNoGo1}) vanishes as well.
Orientifold planes have negative tension, violating the NEC, but as local sources they cannot violate the NEC at every point in the extra dimensions as is required to satisfy (\ref{GKPNoGo1}), as previously discussed.
Aside from warping, then, (which we will return to shortly), there are no ingredients at this step that can evade the constraint

Some interesting recent work has focused on whether the integrated 10-dimensional energy momentum tensor arising from gaugino condensation can evade existing no-go theorems on obtaining 4-dimensional dS \cite{Hamada:2019ack,Carta:2019rhx,Gautason:2019jwq}.
However, gaugino condensation occurs on the worldvolume of D7-branes, which are necessarily constrained to submanifolds within the internal manifold.
Evaluating (\ref{GKPNoGo1}) at points away from these submanifolds, then, means that they cannot contribute to the right-hand side, regardless of the sign of their contribution.
Finally, anti D3-branes have positive tension (although opposite $C_4$ charge), and do not violate the NEC. Further, being local objects, they do not contribute to the right-hand side of (\ref{GKPNoGo1}) at points away from their location.
Thus, we see that none of the ingredients (aside from warping) of the KKLT model can contribute positive terms to the right-hand side of (\ref{GKPNoGo1}) homogeneously throughout the extra dimensions.

It appears, then, that the only way to potentially satisfy (\ref{GKPNoGo1}) for the set of ingredients in KKLT is through the warp factor contributions. The sign and magnitude of the contributions from the warp factor to (\ref{GKPNoGo1}) can be computed based on the first step of KKLT because upon compactification, the fluxes and local sources, such as D3-branes and O3-planes, fix the warp factor.
The ingredients of the later steps, including gaugino condensation and anti D3-branes, are assumed to not perturb the warp factor or internal space so that the 4-dimensional EFT analysis is valid.
The source equation for the warp factor can be computed from the Einstein equations 
(see Appendix \ref{appendix:GKP} for more details)
\be
2 e^{4A} \bar \nabla^2 A = \frac{g_s}{4} e^{8A} \bar{|G_3|}^2 + 8 e^{4A} \bar{(\partial A)}^2 +\frac{7-p}{4} \kappa_{10}^2 T_p e^{2A} \delta(\Sigma)
\label{GKPWarpFactor}
\ee
where contractions in the internal space are made with $\bar g_{mn}$ and the internal space is compact and Ricci flat, $\bar R_{mn} = 0$.
Inserting (\ref{GKPWarpFactor}) into (\ref{GKPNoGo1}), we obtain
\be
3\left(\dot H + H^2\right) =  \frac{g_s}{4} e^{8A} \bar{|G_3|}^2 + 8 e^{4A} \bar{(\partial A)}^2 +\frac{7-p}{4} \kappa_{10}^2 T_p e^{2A} \delta(\Sigma)- 8 \t n^m \t n^n \partial_n A \partial_m A - e^{4A} R_{MN} N^M N^N\, .
\label{GKPNoGo2}
\ee
At first glance, (\ref{GKPNoGo2}) looks promising, since substituting in for the warp factor has provided two positive-definite terms on the right-hand side.
However, there are a number of negative terms on the right-hand side of (\ref{GKPNoGo2}) arising from the last term involving the NEC.
We can use the Einstein equations to calculate the last term (again, see Appendix \ref{appendix:GKP} for details)
\be
R_{MN} N^M N^N &=& \kappa_{10}^2 T_{MN} N^M N^N \nonumber \\
    &=& \frac{g_s}{4} e^{4A} \bar{|G_3|}^2 + 8 \bar{(\partial A)}^2 - 8 e^{-4A} \t n^m \t n^n \partial_n A \partial_m A + \kappa_{10}^2 T_p e^{-2A} \delta(\Sigma) - \kappa_{10}^2 T_p e^{-4A} \Pi_{mn}^\Sigma \t n^m \t n^n \delta(\Sigma)\, .
\label{GKPNEC}
\ee
Substituting (\ref{GKPNEC}) into (\ref{GKPNoGo2}),  the constraint (\ref{GKPNoGo2}) now becomes
\be
3\left(\dot H + H^2\right) = - \frac{p-3}{4} \kappa_{10}^2 T_p e^{2A} \delta(\Sigma) + \kappa_{10}^2 T_p \Pi_{mn}^{\Sigma} \t n^m \t n^n \delta(\Sigma)\, .
\label{GKPNoGo3}
\ee
where the previously positive terms on the right-hand side arising from warping have now cancelled against negative terms arising from the NEC.

The remaining contributions on the right-hand side of (\ref{GKPNoGo3})
are all local sources, and so cannot solve (\ref{GKPNoGo3}) in a homogeneous way.
More generally, (\ref{GKPNoGo3}) illustrates the difficulty in using local sources to ``uplift'' to a dS solution: the left-hand side of (\ref{GKPNoGo3}) is independent of the internal coordinates, while the right-hand side contains delta-functions. If (\ref{GKPNoGo3}) were integrated over the internal space, it might appear possible to obtain a dS space by including appropriate local sources that violate the NEC. However, this expectation would be misleading, since the localized version of the constraint cannot be solved.

We see, then, that none of the ingredients of the KKLT model can provide a homogeneous positive term on the right-hand side of (\ref{GKPNoGo1}); while local violations of the NEC exist in this model (primarily through orientifold planes, and potentially through gaugino condensation as well \cite{Carta:2019rhx,Gautason:2019jwq,Hamada:2019ack}), they cannot solve the constraint at every point in the extra dimensions.

An important part of this argument is that the warp factor and complex structure moduli are fixed and stabilized by the GKP ingredients so that the use of the corresponding 4-dimensional EFT is valid.
Thus, the warp factor cannot respond to the presence of the gaugino condensates or uplifting anti D3-branes.
Perhaps this assumption that the complex structure moduli and warp factor remain fixed is not valid. 
While this would cast doubt on the 4-dimensional EFT analysis, it might provide a path toward understanding how to solve the constraint (\ref{GKPNoGo1}).
However, this would imply that the warp factor must now necessarily include some non-trivial dependence on the cosmological constant.
For example, the anti D3-brane responsible for uplifting is often found at the tip of a strongly warped ``throat'' region.
A common example of such a throat region is the Klebanov-Strassler (KS) throat \cite{Klebanov:2000hb}, supported by $H_3$ and $F_3$ flux.
At the tip of the KS throat, the warp factor approaches a constant $A\approx A_0$ (which determines the maximum redshift of the throat region) so that $\bar \nabla^2 A \approx \partial A \approx 0$.
However, if the warp factor now contains some dependence on the cosmological constant, one of the terms in (\ref{GKPNoGo1}) must be proportional to $H^2$.
For example, taking $\partial_n A \sim H$ implies that the warp factor at the tip of the KS throat is no longer a constant, but instead $e^{A} \sim e^{A_0+H y}$, which could have interesting implications on the dynamics of branes and anti-branes in cosmology.

More broadly, it would be interesting to extend this analysis to flux backgrounds with warping and parallel D$p$/O$p$ sources, such as those in \cite{Andriot:2016xvq,Andriot:2016ufg,Andriot:2018ept} or the dS solutions of \cite{Cordova} (see also \cite{Cribiori:2019clo}).
We leave investigations of these interesting directions for future work.

\subsubsection*{Higher Curvature Corrections}

We have been focusing on the warp factor and energy-momentum contributions to the constraints (\ref{exNoGo2}),(\ref{GeneralNoGo}), however corrections to the Einstein equations in the form
\be
R_{MN} - \frac{1}{2} g_{MN} R_D = \kappa_D^2 T_{MN} + \frac{1}{2}\lambda_{(n)} H_{MN}^{(n)}
\ee
can provide additional terms to the right-hand side of the constraints through
\be
R_{MN} N^M N^N = \kappa_{D}^2 T_{MN} N^M N^N + \frac{1}{2} \lambda_{(n)} H_{MN}^{(n)} N^M N^N\, ,
\label{HigherCurv}
\ee
In string theory models, for which corrections of the form (\ref{HigherCurv}) can be calculated, we have $\lambda_{(n)} \sim \alpha'^{n-1} \sim \ell_s^{2(n-1)}$ in terms of the string length $\ell_s$.
Computing the correction terms in (\ref{HigherCurv}) can be challenging for even the simplest backgrounds; for example, \cite{Burger:2018hpz} computed the corrections arising from Gauss-Bonnet corrections in simple black hole and cosmological backgrounds. We will take the corrections to scale as some power of the curvature $H_{MN}^{(n)}N^M N^N \sim -[R]^n$, and we will assume that these corrections maximally violate the NEC in order to arrive at the strongest possible case for the role of these corrections, so that (\ref{GeneralNoGo}) becomes:
\be
3 H^2 \sim \ell_s^{2(n-1)} [R]^n + ...
\label{CorrectionNoGo}
\ee
where $+...$ denote the other, non-positive contributions to the right-hand side.

We expect the curvature corrections to be dominated by either the 4-dimensional curvature scale $H_{MN}^{(n)}N^M N^N \sim [R]^n \sim H^{2n}$ or the curvature scale of the extra dimensions $H_{MN}^{(n)}N^M N^N \sim [R]^n \sim 1/L^{2n}$. If the former is true, then in order to get dS, the NEC violation must be of the same order as the left-hand side of (\ref{CorrectionNoGo}) so that 
\be
H^2 \sim \ell_s^{2(n-1)} H^{2n}, \hspace{.4in} \Rightarrow \hspace{.4in} \left(H\ell_s\right)^{2(n-1)} \sim 1
\ee
implying that the expansion rate must be the same order as the string scale $H \sim \ell_s^{-1}$, much too large for cosmological interest.
Alternatively, if the curvature scale is dominated by the extra dimensions, then (\ref{CorrectionNoGo}) implies that 
\be
H^2 \sim \frac{\ell_s^{2(n-1)}}{L^{2n}},\hspace{.4in}\Rightarrow \hspace{.4in} L \sim \ell_s\frac{1}{\left(\ell_s H\right)^{1/2n}}
\ee
Interestingly, given the current value of accelerated expansion $H\sim 10^{-41}$ GeV, and taking $\ell_s \sim \left(10^{18} \mbox{ GeV}\right)^{-1}$, this corresponds to a curvature scale of $L \sim 10^{15} \ell_s \sim 1 \mbox{ TeV}$ for $n=6$.

Alternatively, it is possible that the NEC-violating contributions from the higher order corrections to Einstein equations cancel against other NEC-satisfying terms arising from the matter content, so that
\be
H^2 \sim -R_{MN}N^M N^N + ... \sim -\kappa_D^2 T_{MN}N^M N^N +\frac{1}{2}\lambda_{(n)} H_{MN}^{(n)}N^M N^N + ... \sim -\frac{1}{L^2} + \ell_s^{2(n-1)}\frac{1}{L^{2n}}\, .
\ee
However, since $H^2 \ll 1/L^2$, we must then have $L \sim \ell_s$ in order for the correction terms to cancel the negative contributions from the matter content, so that it is not possible to have solutions with parametrically large volume.

The difficulty of realizing positive terms on the right-hand side of the constraint (\ref{GeneralNoGo}) due to higher curvature corrections such as (\ref{HigherCurv}) may have implications for realizing the Large Volume Scenario \cite{Balasubramanian:2005zx,Conlon:2005ki}
as an explicit solution in 10-dimensions.

\subsection{Summary}

We have shown how combining the null Raychaudhuri equation with a fairly generic metric ansatz for a spacetime with static extra dimensions and 4-dimensional dS leads to a simple constraint (\ref{GeneralNoGo}) between the internal curvature, warping, and matter content of the model.
In particular, we have shown that if a model does not have positive curvature,  warping with curvature of the order of the 4-dimensional dS scale, or violate the NEC pointwise everywhere in the bulk, then it cannot support 4-dimensional dS space.

We demonstrated with some explicit examples that the constraint (\ref{GeneralNoGo}) is stronger than existing no-go theorems, and relies on far fewer assumptions. In particular, our constraint is independent of the form of the Einstein equations, matter content, and does not depend on integrals over the extra dimensions.
We demonstrated two simple examples that evade our constraint: a Freund-Rubin compactification of 2-form flux on an $S^2$ with a bulk cosmological constant and an RS model in which the brane-localized and bulk cosmological constants are detuned. Finally, we discussed the role that warping can play, particularly in RS and flux compactification models, and roughly sketched some  difficulties to using higher order corrections to Einstein's equations to evade the constraint.

\section{Apparent Horizons in the Extra Dimensions}
\label{sec:Horizon}

In the previous section we derived a set of necessary conditions to obtain 4-dimensional dS space with extra dimensions, starting from the metric (\ref{exMetric}) (or more generally (\ref{GeneralMetric})).
In this section, we will demonstrate that these cosmological metrics have a surprising consequence: the existence of an anti-trapped horizon in the extra dimensions.
To demonstrate this, we will examine in more detail the properties of the expansion $\theta$ of null congruences.

\subsection{Example: An $S^2$}

We start with the simple case where the spatial metric consists of a 2-sphere, as in our example from Section (\ref{sec:FR}): 
\be
ds^2 = -dt^2 + R^2 \left(d\chi^2 + \sin^2\chi\, d\alpha^2\right)
\label{S2Metric}
\ee
with null rays pointing in opposite directions along $\chi$:
\be
N_{\pm}^M &=& \underset{t,\hspace{.1in}\chi,\hspace{.1in}\alpha}{\left(1, \pm \frac{1}{R},0\right)}\, . \label{S2NullRay}
\ee
It is straightforward to check that these null rays are (affine) geodesics, satisfying $N^M\nabla_M N^N = 0$. The tangent vector $N_+^M$ corresponds to null rays starting at the north pole and converging at the south pole, while the tangent vector $N_{-}^M$ is the opposite, as in Figure \ref{fig:S2NullRays}. 
Also note that the while the component of the tangent vector in the $\chi$-direction $N^\chi_+$ is always positive, the vector is only outgoing for $0 < \chi < \pi/2$, while for $\pi/2 < \chi < \pi$ it is ingoing. This expectation is confirmed by the behavior of the so-called ``areal radius'' $\t r = R \sin\chi$: the rate of change of the areal radius with the $\chi$-coordinate $\frac{d}{d\chi} \t r = R \cos\chi$ is positive for $0 < \chi < \pi/2$, corresponding to outgoing rays, while it is negative for $\pi/2 < \chi < \pi$, corresponding to ingoing rays.

\begin{figure}[t]
\centering\includegraphics[width=.3\textwidth]{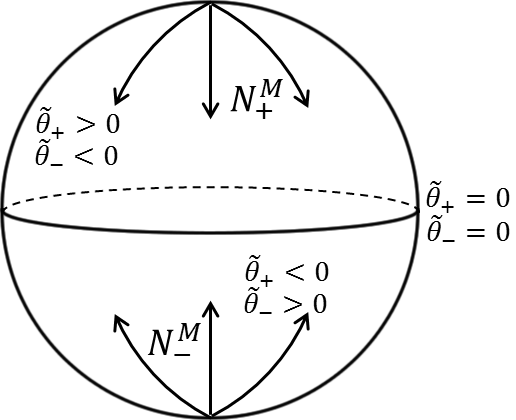}
\caption{Null rays $N^M_\pm$ (\ref{S2NullRay}) diverge at the north or south poles of the $S^2$ and converge at the opposite pole. The expansion of $N^M_+$ is positive in the northern hemisphere and negative in the southern hemisphere, with opposite statements for $N^M_-$.}
\label{fig:S2NullRays}
\end{figure}

The expansion of this congruence of null rays is
\be
\t \theta_{\pm} \equiv \frac{1}{\sqrt{-g_3}} \partial_A \left(\sqrt{-g_3} N^A_\pm \right) = \pm \frac{\cot\chi}{R}\, .
\label{S2Expansion}
\ee
We can see that $\t \theta_{\pm}\rightarrow \pm \infty$ as $\chi \rightarrow 0$, confirming that $N_{\pm}^M$ converge/diverge there (and similarly for $\chi \rightarrow \pi$), while the expansion vanishes $\t \theta_{\pm} = 0$ at the equator $\chi = \pi/2$.
For $0 < \chi < \pi/2$, the expansion is positive (negative) for $N_+^M$ ($N_-^M$), in agreement with our conclusion that the rays are outgoing (ingoing) in this region. Similar statements apply in the lower hemisphere $\pi/2 < \chi < \pi$.
Note that even though the expansions vanish at the equator, this does not mark the location of an apparent horizon; as reviewed in Appendix \ref{appendix:Horizons}, the existence of an apparent horizon requires at least one expansion to be non-zero.

The integral of the expansion over the $S^2$ vanishes for both types of null rays
\be
\int\sqrt{g_2}\, \t\theta_{\pm} d\chi d\alpha = \int_0^{2\pi}\int_0^\pi \pm R \cos\chi\ d\chi d\alpha =  0\, ,
\ee
implying that there is equal amounts of divergence and convergence over the sphere. The vanishing of the integral of the expansion is due simply to the fact that the expansion is a total derivative on the sphere
\be
\sqrt{g_2}\, \t \theta_{\pm} = \partial_m \left(\sqrt{g_2} N_\pm^m\right)\, ,
\ee
where we used $\sqrt{-g_3} = \sqrt{g_2}$ from the metric (\ref{S2Metric}). The vanishing of the expansion when integrated over the compact space will be important when we consider the more general case: it implies that there are equal amounts of positive and negative expansion, and thus there must be a locus upon which the expansion vanishes $\t \theta_{\pm} = 0$.
Thus, this is a generic feature not restricted to the $S^2$.

Now, let's consider the $S^2$ as a direct product with an expanding 4-dimensional FLRW space\footnote{We are considering the 3-dimensional spatial metric to be flat for simplicity, though the results hold for isotropic and homogeneous spatial metrics with positive and negative curvature as well.} with scale factor $a(t)$
\be
ds^2 = -dt^2 + a^2(t)\delta_{ij} dx^i dx^j + R^2\left(d\chi^2 + \sin^2\chi d\alpha^2\right)\, .
\label{S2FRWMetric}
\ee
As with the previous example, the detailed matter content giving rise to this metric is unimportant; we are only concerned with the geometric properties of the metric, supposing that it exists.
For our purposes, we choose a pair of affine null tangent vectors with one leg along the $\chi$ direction
\be
N_\pm^M = \underset{t,\hspace{.1in}\vec{x},\hspace{.1in}\chi,\hspace{.1in}\alpha}{\left(1,\ \vec{0},\ \pm \frac{1}{R},\ 0\right)}\, .
\label{S2NullRayFRW}
\ee
Note that the affine null vectors (\ref{S2NullRayFRW}) do not contain any time-dependence or functional dependence on the scale factor $a(t)$: 
their interpretation is the same as before, as outgoing and ingoing in their respective hemispheres.

\begin{figure}[t]
\centering\includegraphics[width=.8\textwidth]{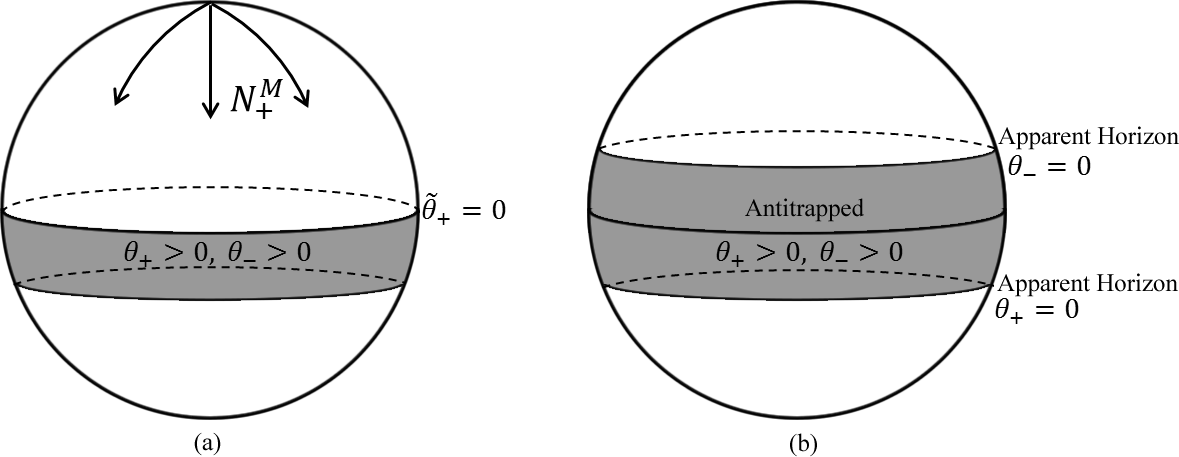}
\caption{(a) The expansion $\theta_+$ of the null rays $N^M_+$ is positive not only in the northern hemisphere, but also in a band in the southern hemisphere because the expansion of the congruence of null rays also depends on the Hubble expansion, as in (\ref{S2ExpansionFRW}). This band in the southern hemisphere is antitrapped because both expansions are positive in this region.
(b) Repeating the same argument for $N^M_-$, we see that there is a band on both sides of the equator of antitrapped spacetime in which the expansions of oppositely oriented null vectors are both positive. This antitrapped spacetime is bounded by apparent horizons.}
\label{fig:S2Horizons}
\end{figure}

The expansion for the null rays (\ref{S2NullRayFRW}) depends not only on $\chi$, but also on the non-compact scale factor $a(t)$ through the determinant of the 6-dimensional metric
\be
\theta_{\pm} = \frac{1}{\sqrt{g_6}} \partial_A \left(\sqrt{g_6} N^A_\pm \right) = \frac{1}{a^3 R^3 \sin\chi} \partial_A \left( a^3 R^3 \sin\chi N^A_\pm\right) = 3 H \pm \frac{\cot\chi}{R}\, ,
\label{S2ExpansionFRW}
\ee
where $H \equiv \frac{\dot a}{a}$, with a dot denoting a derivative with respect to time, and the last term is $\t \theta_\pm$ as in (\ref{S2Expansion}).
Interestingly there is now a band, with width dependent on the 3-dimensional expansion rate $H$, in which both null tangent vectors have positive expansion. 
To see this, consider the null vector outgoing from the north pole $N_+^M$. Certainly this vector has positive expansion $\theta_+$ up to the equator for $0 < \chi < \pi/2$ as before, as in Figure \ref{fig:S2Horizons}. However, due to the presence of the Hubble expansion term the expansion is also positive beyond the equator up to the angle $\chi < \pi/2 + \cot^{-1}(3 H R)$, reaching into the lower hemisphere where this null ray is ingoing. But note that the expansion $\theta_-$ for the null rays outgoing from the south pole $N_-^M$ is also positive in this region. Thus, we see that in the region $\pi/2 < \chi < \pi/2 + \cot^{-1}(3 H R)$ both ingoing and outgoing null rays have positive expansion. A similar argument 
follows for the region $\pi/2 - \cot^{-1} (3 H R) < \chi < \pi$, in which we see that both expansions are positive in this region as well.

Altogether, we have found a band around the equator $\pi/2 - \cot^{-1} (3 H R) < \chi \pi/2 + \cot^{-1}(3 H R)$ in which both expansions are positive, so that this region of spacetime is {\it antitrapped}, and the boundaries are {\it apparent horizons} (see Appendix \ref{appendix:Horizons}).
Further, these apparent horizons are (inner past) {\it trapping horizons} when ${\mathcal L}_+ \theta_- > 0$, which occurs when
\be
{\mathcal L}_+ \theta_- = N_+^M \partial_M \left(3 H - \frac{\cot \chi}{R}\right) = 3 \dot H +\frac{\csc^2\chi}{R^2} = 3\dot H + \frac{1+(3HR)^2}{R^2} > 0\, .
\ee
Note that for de Sitter space $\dot H = 0$ and near-de Sitter space $\dot H = -\epsilon H^2$ (for $\epsilon \ll 1$), this condition is satisfied automatically. Indeed, the condition for the existence of trapping horizons is satisfied up until the rate of change of the Hubble expansion is of the order of the size of the $S^2$, $\dot H \sim -\frac{1}{R^2}$.

In order to recover effective 4-dimensional physics, we expect that the Hubble expansion rate is much smaller than the size of the radius of the $S^2$, $H R \ll 1$, so that the antitrapped band is quite thin compared to the size of the $S^2$.
However, if the Hubble scale were to become large compared to the size of the sphere (as in the early Universe), then nearly the entire $S^2$ would be encompassed by the antitrapped region.

\subsection{General Warped Case}

Let us now consider a more general $D$-dimensional metric consisting of a warped product between a macroscopic expanding 3-dimensional spacetime and a $n$-dimensional compact internal space, of the form (\ref{GeneralMetric})
\be
ds_D^2 = \Omega^2(y^m)\left[-dt^2 + a^2(t) \delta_{ij} dx^i dx^j + \t g_{mn}(y^m) dy^m dy^n\right]\, .
\label{GeneralHorizonMetric}
\ee
As before, we will only be concerned with the geometric properties of this metric ansatz.
Oppositely oriented affine null rays in $D$-dimensions with a leg along the extra dimensions can be written
\be
N^M_{\pm} = \Omega^{-2} \underset{t,\hspace{.1in} \vec{x}, \hspace{.1in} \vec{y}}{\left(1,\ \vec{0},\ \t n_{\pm}^m\right)}
\label{GeneralNullRay1}
\ee
where $\t n_\pm^m \t n_\pm^n \t g_{mn} = +1$ is a unit vector and is affine on the internal metric $\t n^m_\pm \t \nabla_m \t n_\pm^n = 0$.

\begin{figure}[t]
\centering\includegraphics[width=.25\textwidth]{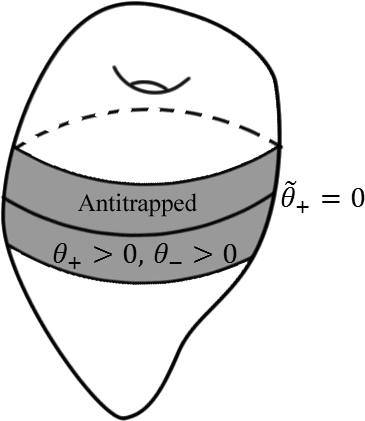}
\caption{In a general compact space, there is a band of antitrapped space around $\t \theta_\pm = 0$ in which the expansions of oppositely oriented null rays are both positive.}
\label{fig:GeneralHorizon}
\end{figure}

The null expansion is
\be
\theta_\pm = \frac{1}{\sqrt{-g_D}} \partial_M \left(\sqrt{-g_D} N^M\right) = 3 H \Omega^{-2} + \t \theta_{\pm}
\label{GeneralExpansion}
\ee
where we defined the expansion on the internal space as
\be
\t \theta_\pm \equiv \frac{1}{\Omega^D \sqrt{\t g_n}} \partial_m \left(\Omega^{D-2} \sqrt{\t g_n} \t n^m_\pm\right)\, .
\label{InternalExpansion}
\ee
The expansion on the internal space is proportional to a total derivative on the compact space, and so its integral vanishes
\be
\int \sqrt{\t g_n}\, \Omega^D\, \t \theta_\pm\, d^ny = 0\, .
\label{GeneralIntegral}
\ee
As in the case of the $S^2$, this implies that there are equal amounts of positive and negative $\t \theta_\pm$ expansion throughout the internal space, and thus there must be a locus on which $\t \theta_\pm = 0$. We can intuitively understand this locus as the submanifold upon which the null rays (\ref{GeneralNullRay1}) transition from ingoing to outgoing. 

As in the case of the $S^2$, there is now a band of {\it antitrapped spacetime} surrounding the locus upon which $\theta_\pm = 0$, as in Figure \ref{fig:GeneralHorizon}. To see this, note that for an arbitrarily small distance from the locus where $\t\theta_\pm = 0$ in one direction there is a region where $\t \theta_-$ is negative and arbitrarily small, while $\t \theta_+$ is positive (corresponding to $N_-^M$ ingoing and $N_+^M$ outgoing). However, since $H> 0$, this implies that $\theta_- = 3 H \Omega^{-2} + \t \theta_- > 0$ is positive; since $\theta_+$ is still positive in this region, we have a region of spacetime where both $\theta_+, \theta_-$ are positive, signaling the existence of an antitrapped spacetime. Moving further from $\t \theta_\pm = 0$, we eventually reach a point where $\theta_- = 0$ and $\theta_+ > 0$, marking the location of an apparent horizon.
A similar argument moving away from $\t \theta_\pm = 0$ in the opposite direction leads to another antitrapped region and apparent horizon.
(Unfortunately, it is not possible to determine conclusively that these boundaries are also trapping horizons given the level of generality of the metric.)

\subsection{Discussion}

We have shown, through the simplified example of a direct product of a 3-dimensional spacetime with an $S^2$ as well as that of a more general warped product, that the existence of antitrapped regions and apparent horizons is a generic feature of 4-dimensional cosmological spacetimes with stabilized extra dimensions. Before we discuss the possible implications of these results, we should briefly remark on the assumptions made and their limitations.

First, the generic metric ansatz (\ref{GeneralHorizonMetric}) assumes that the internal dimensions are static, so that the size of the extra dimensions is stabilized. If the internal metric $\t g_{mn}$ is time-dependent, this gives rise to additional terms in the expansion (\ref{GeneralExpansion}) which will affect the location and possibly even the existence of the antitrapped region.
We have also assumed that, aside from the warping factor $\Omega(y)$, the $D$-dimensional metric is a direct product between the 3-dimensional and internal space. Unless the 4-dimensional spacetime is de Sitter, it is possible to also allow for an off-diagonal mixing term in the metric $g_{\mu m}(t,y^m)$. Such a term can be removed by an appropriate $D$-dimensional diffeomorphism, but at the expense of inserting time-dependence into the internal metric or additional $y$-dependence into the 4-dimensional metric beyond that of the warp factor. For simplicity we will not consider such terms, though it would be interesting to see how they would change our conclusions.

Finally, we have assumed that the internal space is compact so that the integral (\ref{GeneralIntegral}) vanishes, thus implying the existence of a locus upon which $\theta_{\pm} = 0$. This integral does not automatically vanish for spacetimes that are bounded by branes (such as RS \cite{RS2} or Football geometries \cite{Aghababaie:2002be,Carroll:2003db,Aghababaie:2003wz,Tolley:2005nu}); however, if these spacetimes nonetheless have a locus upon which $\theta_{\pm} = 0$ then we still expect there to be antitrapped regions in the extra dimensions.

Let us now turn to the interpretation of these results. It is puzzling to find the presence of an antitrapped region in the extra dimensions of (\ref{GeneralIntegral}), since the extra dimensions themselves are static and not participating in the expansion.
The origin of the antitrapped region can be seen by the presence of {\it shear}. The null rays (\ref{GeneralNullRay1}) have a vanishing tangent vector component along the 3-dimensional comoving coordinates $N_\pm^i = 0$. This implies that these trajectories are at fixed {\it comoving} coordinates $x^i$, but expanding {\it physical} coordinates $X_{phys} = a(t) x^i$.
Thus, a congruence of nearby null rays, which are nearby not only in $y^m$ but also in $x^i$, expands due to the expansion of the 3-dimensional space as the rays are ``pulled along'' with the comoving coordinates.
Note that
since the expansion $\theta_{\pm}$ is a scalar, it is not possible to decompose it into, say, shear and parallel expansion, arising from the expanding 3-dimensions and extra dimensions, respectively.
Thus there does not seem to be a generally covariant way to distinguish between apparent horizons arising due to the direct expansion of coordinates and those due to shear of transverse coordinates.

Despite the existence of an apparent horizon, however, it appears that a null ray can traverse from one end of the extra dimensions to the other, and in particular can traverse across the apparent horizon, in finite affine time, as long as $\Omega$ does not contain singularities, as can be seen by integrating  (\ref{GeneralNullRay1}).
Thus, it is not immediately clear what the physical significance of the antitrapped region is for the cosmology of metrics with extra dimensions. 
Perhaps bulk modes that are spread out throughout the extra dimensions can feel the effects of the apparent horizon over timescales comparable to the Hubble time.
We hope to pursue this further in the future.

\section*{Acknowledgements}

This work was supported by the
Natural Sciences and 
Engineering Research Council of Canada. 

\appendix

\section{GKP Backgrounds}
\label{appendix:GKP}

A commonly used class of backgrounds utilize the low-energy limit of IIB string theory, with the action
\be
S_{\rm IIB} = &&\frac{1}{2\kappa_{10}^2} \int d^{10}x \sqrt{-g} \left({\mathcal R} - \frac{\partial_M \tau \partial^M \bar \tau}{2 ({\rm Im} \tau)^2}\right) - \frac{1}{2\kappa_{10}^2}\int \left[\frac{G_3\wedge \star \bar G_3}{12\,{\rm Im}\tau} + \frac{1}{4} \t F_t \wedge \star \t F_5 \right. \nonumber \\
&& \left. + \frac{i}{4{\rm Im}\tau} C_4\wedge G_3 \wedge \bar G_3\right] + S_{\rm loc}\, ,
\ee
where ${\mathcal R}$ is the 10-dimensional Ricci scalar, $\tau = C_0 + ie^{-\phi}$ is the axio-dilaton, the 3-form gauge field strengths $F_3 = dC_2$ and $H_3 = dB_2$ have been combined into the complex 3-form $G_3 = F_3 - \tau H_3$, and we defined the five-form field strength as $\t F_5 = dC_4 - C_2 \wedge H_3$. The action for local objects, such as D3-branes and O3-planes, is included in $S_{\rm loc}$.

Our background will consist of the metric and fluxes \cite{Giddings:2001yu}
\be
&&ds_{10}^2 = e^{2A(y^m)} \hat \eta_{\mu\nu} dx^\mu dx^\nu + e^{-2A(y^m)} \bar g_{mn} dy^m dy^n\, ; \nonumber \\
&&\t F_5 = \hat \epsilon_4 \wedge \bar d_6 e^{4A} + \bar \star d_6 e^{-4A}\, ; \hspace{.2in} \bar \star_6 G_3 = i G_3\, ;
\ee
with a constant $\tau = i g_s$.
The Ricci tensor components are
\be
R_{\mu\nu} &=& -\hat\eta_{\mu\nu} e^{4A} \bar \nabla^2 A\, ; \\
R_{mn} &=& \bar R_{mn} + \bar g_{mn} \bar \nabla^2 A - 8 \partial_n A \partial_m A\, .
\ee
The contributions from the fluxes and localized sources lead to the following energy momentum tensors:
\be
T_{\mu\nu}^{(3)} &=& -\frac{g_s}{4\kappa_{10}^2} e^{8A}\bar{|G_3|}^2 \hat \eta_{\mu\nu} \\
T_{mn}^{(3)} &=& \frac{g_s}{4\kappa_{10}^2} \left(e^{4A} G_m^{\bar p \bar q} \bar G_{npq} - \bar g_{mn} e^{4A} \bar{|G_3|}^2\right) \\
T_{\mu\nu}^{(5)} &=&-\frac{4}{\kappa_{10}^2} e^{4A} \bar{(\partial A)}^2 \hat \eta_{\mu\nu} \\
T_{mn}^{(5)} &=& \frac{1}{\kappa_{10}^2} \left[4 \bar{(\partial A)}^2 \bar g_{mn} - 8 \partial_m A \partial_n A\right] \\
T_{\mu\nu}^{\rm loc} &=& -T_p \hat \eta_{\mu\nu} e^{2A} \delta(\Sigma) \\
T_{mn}^{\rm loc} &=& -T_p \Pi_{mn}^{\Sigma} \delta(\Sigma)
\ee
where we have denoted the submanifold of the localized source as $\Sigma$, and $\Pi_{mn}^\Sigma$ is the projector onto this submanifold, with $\Pi_{mn}^\Sigma g^{mn} = (p-3)$.

The equation for the warp factor can be obtained by tracing over the 4-dimensional components of the trace-revered Einstein equation $R_{\mu\nu} = \kappa_{10}^2 \left(T_{\mu\nu} - \frac{1}{8} g_{\mu\nu} T^M_M\right)$, giving the following equation
\be
2 e^{4A} \bar \nabla^2 A = \frac{g_s}{4} e^{8A} \bar{|G_3|}^2 + 8 \bar{(\partial A)}^2 e^{4A} - \frac{(p-7)}{4} \kappa_{10}^2 T_p e^{2A} \delta(\Sigma)\, .
\ee

\section{Apparent and Trapping Horizons}
\label{appendix:Horizons}

Consider a pair of oppositely oriented affine null rays $n_\pm^M$ (``ingoing'' $(-)$ and ``outgoing'' $(+)$) with corresponding expansions $\theta_\pm$.
Following \cite{PhysRevD.49.6467} (see also \cite{Faraoni:2011hf}), an {\it apparent horizon}  is a surface on which one of the expansions vanishes while the other is non-zero. 
A region of spacetime in which both expansions are positive is called {\it antitrapped}, while a region of spacetime in which both expansions are negative is called {\it trapped}.

For example, the apparent horizon of a Schwarzschild black hole occurs where the expansion for ingoing null rays is negative $\theta_- < 0$  while the expansion for outgoing null rays vanishes $\theta_+ = 0$. The region inside of this apparent horizon, where both expansions are negative $\theta_+, \theta_- < 0$ is a trapped region. Similarly, a cosmological apparent horizon in 4-dimensional spacetime occurs when the the expansion of the outgoing null ray $\theta_+$ is positive, while the expansion for the ingoing null ray vanishes $\theta_- = 0$. The region outside of this apparent horizon, where both expansions are positive $\theta_+, \theta_- > 0$ is an antitrapped region.

Finally, we define a (inner past) {\it trapping horizon} as a surface upon which the Lie derivative of the inward-pointing expansion along the outward pointing null ray is positive \cite{PhysRevD.49.6467}
\be
{\mathcal L}_+ \theta_- > 0\, .
\ee
The trapping horizon provides a better definition for the existence of the boundary of a trapped region than the mere presence of an apparent horizon, as discussed in \cite{PhysRevD.49.6467}.

\bibliographystyle{utphysmodb}

\bibliography{refs}

\end{document}